\documentclass[conference]{IEEEtran}
\IEEEoverridecommandlockouts

\usepackage{float}

\def\BibTeX{{\rm B\kern-.05em{\sc i\kern-.025em b}\kern-.08em
    T\kern-.1667em\lower.7ex\hbox{E}\kern-.125emX}}

\usepackage{cite}
\usepackage{amsmath,amssymb,amsfonts}
\usepackage{algorithmic}
\usepackage{graphicx}
\usepackage{tabularx}
\usepackage{textcomp}
\usepackage{booktabs}
\usepackage{dcolumn}
\usepackage{xcolor}
\usepackage{array}
\usepackage{subfigure}
\newcolumntype{d}{D{.}{.}{-1}}
\newcolumntype{Y}{>{\raggedright\arraybackslash}X}
\newcolumntype{Z}{>{\centering\arraybackslash}X}

\begin{document}

\title{Investment Ranking Challenge : Identifying the best performing stocks based on their semi-annual returns}
\author{\IEEEauthorblockN{Shanka Subhra Mondal}
\IEEEauthorblockA{\textit{Indian Institute of Technology,  Kharagpur}\\
India \\
shankasubhra@iitkgp.ac.in}
\and 
\IEEEauthorblockN{Sharada Prasanna Mohanty}
\IEEEauthorblockA{
\textit{Ecole Polytechnique F\'ed\'erale de Lausanne}\\
Switzerland \\
sharada.mohanty@epfl.ch}
\and
\IEEEauthorblockN{Mehmet Koseoglu}
\IEEEauthorblockA{\textit{University of California}\\
Los Angeles, USA \\
mkoseoglu@ucla.edu}
\and
\IEEEauthorblockN{Lance Rane}
\IEEEauthorblockA{\textit{Imperial College London}\\
United Kingdom \\
lance.rane14@imperial.ac.uk}
\and

\IEEEauthorblockN{Kirill Romanov}
\IEEEauthorblockA{\textit{ }\\
Russia \\
kirill.v.romanov@gmail.com}
\and
\IEEEauthorblockN{Wei-Kai Liu}
\IEEEauthorblockA{\textit{}\\
Taipei, Taiwan, ROC \\
alphard.liu@gmail.com}
\and
\IEEEauthorblockN{Pranoot Hatwar}
\IEEEauthorblockA{\textit{}\\
Mumbai, India \\
pphatwar1995@gmail.com}
\and 
\IEEEauthorblockN{Marcel Salathe}
\IEEEauthorblockA{
\textit{Ecole Polytechnique F\'ed\'erale de Lausanne}\\
Switzerland \\
marcel.salathe@epfl.ch}
\and 
\IEEEauthorblockN{Joe Byrum}
\IEEEauthorblockA{\textit{Principal}\\
City, USA \\
byrum.joe@principal.com}
}

\maketitle

\begin{abstract}
In the IEEE Investment ranking challenge 2018, participants were asked to build a model which would identify the best performing stocks based on their returns over a forward six months window. Anonymized financial predictors and semi-annual returns were provided for a group of anonymized stocks from 1996 to 2017, which were divided into 42 non-overlapping six months period. The second half of 2017 was used as an out-of-sample test of the model's performance. Metrics used were Spearman's Rank Correlation Coefficient and Normalized Discounted Cumulative Gain (NDCG) of the top 20\% of a model's predicted rankings. The top six participants were invited to describe their approach. The solutions used were varied and were based on selecting a subset of data to train, combination of deep and shallow neural networks, different boosting algorithms, different models with different sets of features, linear support vector machine, combination of convoltional neural network (CNN) and Long short term memory (LSTM).
\end{abstract}


\section{Introduction}
Investment decisions are increasingly data-driven, leveraging changing patterns in environmental and stock-level predictors to gain a performance edge. Predicting stock returns is notoriously difficult on account of a low signal-to-noise ratio in the relationships between predictors and target; for the purposes of guiding investment strategy, it is often sufficient to predict stock rankings rather than absolute returns. The potential gains to the analyst of even slight improvements in the ability to pick out well-performing stocks are large. In this competition, participants were tasked with ranking semi-annual returns for a group of anonymized stocks between 1996 and 2017 on the basis of a set of anonymized financial predictors.  Labelled data (predictors and company returns) were provided for all periods except for the final one, for which only predictors were provided. Participants were required to use a sliding-window procedure to generate distinct models at each time period using only the information available prior to that period. Final evaluation was performed using two metrics: Spearman's Rank Correlation Coefficient and Normalized Discounted Cumulative Gain (NDCG) of the top 20\% of a model's predicted rankings. The quality of predicted rankings over the period 1996 to 2016 and that for the final period were given equal weighting in participants' final scores.

\section{Brief Literature Review}

Various machine learning techniques have been used to anticipate economic and environmental changes which are crucial for successful stock market predictions,  fundamental to the formation of investment strategies by using financial data in the form of time series. \cite{zbikowski2015using} used volume weighted support vector machines \cite{cortes1995support} along with F-score based feature selection to forecast short term trends in stock market. \cite{xiao2013ensemble} proposed a three stage neural network based non-linear weighted ensemble where the first stage was used to generate three base neural network models followed by particle swarm optimization \cite{kennedy2010particle} and the final stage learning used SVM neural network.

A hybrid two stage fusion approach including support vector regression (SVR) \cite{basak2007support} in first stage combined with artificial neural networks, random forest \cite{breiman2001random} and SVR in second was developed by \cite{patel2015predicting} for efficient prediction of future values of stock market index. \cite{ballings2015evaluating} bench-marked ensemble methods like Adaboost \cite{freund1997decision}, Random Forest \cite{breiman2001random} and Kernel Factory \cite{ballings2013kernel} against single classifiers like logistic regression, support vector machines, K-Nearest neighbours, Neural Networks and showed Random Forest is the best algorithm for stock price direction prediction.

Owing to the success of deep learning for various tasks it has also been explored in this domain. For instance \cite{rather2015recurrent} proposed a hybrid model which used predictions from autoregressive moving average model, exponential smoothing model and recurrent neural network. They also used genetic algorithms to find weights for the hybrid model. \cite{fischer2018deep} used Long Short Term Memory Networks \cite{hochreiter1997long} to predict out of the sample directional movements for the constituent stocks of the S\&P 500 which outperformed random forests , logistic regression classifier. \cite{krauss2017deep} proposed an ensemble of deep neural networks, gradient boosted trees and random forests for producing out of the sample returns of the stocks in S\&P 500.

Finally, it is interesting to note the results of a similar competition that ran on the Kaggle\textsuperscript{TM} platform in 2017 \cite{kaggle}, where participants were similarly required to predict the rankings of financial instruments on the basis of anonymised predictors. Simple linear models featured heavily in top competitors' solutions. The final winning solution achieved a low R value of 0.038, emphasizing the difficulty of the ranking task.

\section{Training Subset Selection}
Mehmet Koseoglu
\\

Contrary to the popular approaches in supervised learning where all samples in the training set are used for training, the proposed algorithm uses only a subset of the training data to train the model. In this approach, we try to find the periods which improves prediction accuracy for the target period and include only those periods into the training set.  By searching through the periods in the training set and using the performance of the test data, the algorithm finds the optimum periods to include in the dataset. 

\subsection{Method}
The proposed training subset selection algorithm iteratively searches through the periods to be included in the training set. The algorithm initially starts with the complete training dataset. The supervised learning algorithm is trained using the whole training dataset and a prediction accuracy is obtained. Then, the algorithm removes the first period from the training dataset and re-trains the supervised learning algorithm. If the removal of the first period improves prediction accuracy, that period is kept removed from the training set. The algorithm then moves to the second period, removes it from the training data and evaluates the prediction accuracy. Similarly, the second period is kept removed from the dataset if removal improves accuracy. The algorithm goes through all periods in a similar fashion. After the first pass over all training periods, the algorithm restarts the procedure again. This process is repeated until the training set converges. 

Along with the training subset selection algorithm, we have used the Bayesian linear regression as the supervised learning algorithm. By having a Gaussian prior on the parameters, the model prevents overfitting to the training data. More details about the training subset selection and the supervised learning algorithm can be found in a separate paper \cite{}. 

\subsection{Experiments and results}
Our experiments indicate that the training subset selection improves the prediction accuracy significantly. We have observed that when the whole training data is used, the Spearman’s correlation is around 0 whereas when our training subset selection algorithm is used, we obtain a Spearman’s coefficient of 0.26. 

\subsection{Discussion}
Our results suggest that it is possible to improve prediction accuracy significantly by selectively constructing the training dataset. The main reason behind is that the training data includes periods from different market conditions some of which may not reflect the market conditions of the target period. The training subset selection algorithm implicitly selects the periods which have similar characteristics wih the target period into the training set. 

\section{Deep and shallow methods for asset ranking prediction}
Lance Rane
\\

A combination of ridge regression models and deep neural networks were used to predict rankings of anonymized financial asset returns over 42 non-overlapping periods, using a sliding-window technique. Performance across periods was highly variable but consistently outperformed random stock rankings, with feature selection found to be a particularly significant determinant of performance.

\subsection{Method}
Minimal pre-processing and feature engineering were performed. Missing values were replaced by zeroes, and new features were created by taking the mean of all features across the six months of a given period for which observations were available, increasing the number of available features to 493. 

For all periods bar the final one, labelled data were available to provide validation feedback for model selection procedures. For these periods, separate models were trained for each period using all the data available up to that period – known henceforth as the training set – using a sliding window procedure. Features in the training set were ordered by correlation with the target value and feature selection proceeded by stepwise selection using 60\% of the labelled training data of the current period to provide feedback. A ridge regression model was used to evaluate the utility of specific features and hyperparameters were tuned using the remaining 40\% of the current period data to provide feedback. In other words, candidate models were tested using out-of-sample data, so as to provide an unbiased estimate of final testing performance. 

For the final period, no validation data were available, so to improve the chances of good performance on this relatively small dataset, feature selection was performed by generic validation across all preceding periods. That is, features were selected by simple rules that resulted in good performance across all preceding periods and so were deemed likely to provide reasonable out of sample performance on future test data. This strategy required, once again, the building of models at each period so as to allow for the validation of model selection procedures. 

For training targets of prior period models, normalized return was taken and transformed as follows: 

\begin{itemize}
\item Values were multiplied by a numerical integer index representing the chronological period (ranging between 1 and 42) raised to some power, p.
\item Target values were ranked by magnitude across the entirety of the training dataset.
\item The value, p, was chosen for its cross-validated performance and set at 2.
\end{itemize}

Predictions at prior periods were made using ridge regression models, the hyperparameters of which were chosen for their performance across the validation data. Models used an alpha coefficient of regularization of 850 and excluded intercept parameters. 

For the final period, model selection was performed using the average score across all periods of the validation data and models were not heavily optimized to given periods, to lower the risk of overfitting. Deep feedforward neural networks implemented in Tensorflow were found to perform best. The final model comprised 3 hidden layers of 2000, 1000 and 400 neurons each. ReLU non-linearities were applied throughout except at the final output layer, where a tanh function was used to compress the output to the range (-1,1). Regularisation involved the application of both dropout, after every layer, and weight decay, the parameters of which were tuned for performance across the whole of the validation data. Performance was found to vary with random seed and initialization conditions. Ultimately, the best performing seed found during shallow search using the validation data was employed in the final model.

\subsection{Experiments and results}
For models trained with the benefit of validation feedback, the average calculated spearman coefficient was 0.268. For the model selection procedure applied for the final period, average performance across all previous periods was 0.065. There was significant variation across periods with both methods.

\begin{table}[htbp]
\caption {Model performance: Spearman's Rank} 
\begin{center}
\begin{tabular}{ l c c }
\hline
 & \textbf{avg.} & \textbf{score (s.d.)} \\
\hline
prior periods & 0.268 & (0.132) \\
final periods & 0.065 & (0.102) \\
\hline
\end{tabular}
\label{tab_lance_1}
\end{center}
\end{table}

\subsection{Discussion}
The results suggest a role for complex non-linear methods in the asset ranking problem, where simple models have traditionally been relied upon. The large discrepancy in performance for prior and final period models can be attributed to the more generic methods used for model selection in the final period, in order to reduce the risk of overfitting. 

Transformation by multiplication with a chronological index representing the order of periods was found to be beneficial to cross-validated performance for ridge regression models. By transforming in this way, training data from periods more recent to the test period are scaled more prominently, and thus the range of values from these periods is greater. Values in the tails of these more recent distributions are thus attributed more extreme positions upon subsequent ranking, and thus provide a greater training signal. Intuitively, given the gradual time-varying nature of some market conditions and other indicators relevant to stock value, it would seem that more recent periods might provide more relevant information; this transformation is thought to reflect this. 

\section{Boosted Stock Prediction}
Shanka Subhra Mondal
\\

Portfolio managers need to identify the stocks with extreme positive or negative returns based on the distribution of stock returns. Having the right data at the right time and extracting relevant features from it and then using an appropriate model to fit these features plays a crucial role in successfully anticipating economic and environmental changes that may impact investment performance. My approach broadly consisted of creating new features and using different boosting algorithms to predict stock returns for different time periods based on their validation scores with different sets of features.  
\subsection{Method}
First small amount of feature engineering was done. Each unique variable was broken up into six non-overlapping observations in each time period. For example $X1$ had six monthly observations in each period represented as $X1\_1,X1\_2,...,X1\_6$. To make it easier to model, average of the values within each time period were found out. Also a new feature was created which is the percentile of each variable average. The year and the quarter of the period were also used as feature into our model. Imputation of all missing values were done with zeros. Total $142$ features were created ($70$ average, $70$ average percentile, year and quarter). Expanding window procedure was adopted for training the model up to time $t$ to predict on all observations at time $t+1$. Different models were used for different time periods with different set of features. Four types of regressor models were used namely xgboost \cite{chen2016xgboost}, lightgbm \cite{ke2017lightgbm},random forest \cite{breiman2001random}and catboost \cite{dorogushcatboost}. Lightgbm model (l) used the features average, year, quarter. Lighgbm model (l1) used the features average percentile,year,quarter. Xgboost model(x) used the features similar to lightgbm model(l). Xgboost model (x1) used the features similar to lightgbm model (l1). Random forest used the features average and average percentile only whereas catboost model used all the features. All the parameters used for these models can be found in the code uploaded to gitlab.crowdai.org. The catboost and random forest models were trained only from period 1996\_2 to 2002\_1,wheras others used the expanding window technique. Model along with selected features which gave the highest spearman correlation on training data for next time period was used to predict the test data for that time period.

\subsection{Experiments and results}
The Spearman's correlation obtained in the validation set for the time periods along with the model used are shown on Table~\ref{tab_shanka_1}.

\begin{figure}[htbp]
\caption{Stock Return Prediction on Unseen Data}
\includegraphics[width=0.5\textwidth]{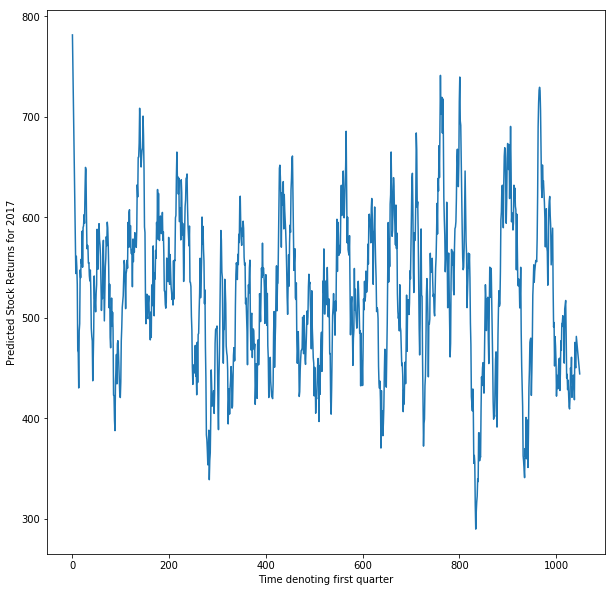}
\label{fig:shanka_result}
\end{figure}

The prediction of the lightgbm model for 2017 first quarter stock returns  is shown in Fig.~\ref{fig:shanka_result}.
\begin{table}[htbp]
\caption{Sperman correlation obtained}
\begin{center}
\begin{tabular}{ l d }
\hline
\textbf{Time Period and model} & \textbf{score} \\
\hline
2002\_1-Lightgbm(l) & 0.117 \\
2002\_2-Xgboost(x) & -0.11 \\
2003\_1-Random forest & 0.058 \\
2003\_2-Catboost & 0.13 \\
2004\_1-Random forest & 0.154 \\
2004\_2-Catboost & 0.077 \\
2005\_1-Lightgbm(l) & 0.08 \\
2005\_2-Lightgbm(l1) & 0.05 \\
2006\_1-Catboost & 0.04 \\
2006\_2-Catboost & -0.03 \\
2007\_1-Lightgbm(l) & 0.125 \\
2007\_2-Lightgbm(l1) & 0.178 \\
2008\_1-Catboost & 0.19 \\
2008\_2-Xgboost(x1) & 0.189 \\
2009\_1-Lightgbm(l) & 0.169 \\
2009\_2-Catboost & -0.007 \\
2010\_1-Lightgbm(l1) & 0.0229 \\
2010\_2-Lightgbm(l1) & 0.064 \\
2011\_1-Lightgbm(l) & 0.134 \\
2011\_2-Random forest & -0.016 \\
2012\_1-Random forest & 0.044 \\
2012\_2-Xgboost(x1) & 0.1216 \\
2013\_1-Random forest & 0.179 \\
2013\_2-Catboost & -0.06 \\
2014\_1-Catboost & 0.209 \\
2014\_2-Lightgbm(l1) & 0.06 \\
2015\_1-Xgboost(x1) & 0.209 \\
2015\_2-Rndom forest & 0.0848 \\
2016\_1-Random forest & 0.143 \\
2016\_2-Catboost & 0.016 \\
\end{tabular}
\label{tab_shanka_1}
\end{center}
\end{table}

\subsection{Discussion}
The results show that predictive modelling with appropriate feature set can be effective in solving the problem addressed in the paper. However the main challenge was to select the model and also what features should be used for prediction of stock rank returns for 2017 time period where there was no validation data to test the model for that period. Hence picked up lightgbm model (l) and its associated features like average,year,quarter because of its faster training speed, low memory usage, higher efficiency. As a part of the future work a proper selection of subset of the training period along with model and feature selection can be tried out which can yield better results.

\section{Kirill Part}
Kirill Romanov
\\

The main goal of the competition was to develop a model that will help identify the best-performing stocks in each time-period using the provided data sets of financial predictors and semi-annual returns. My strategy was the creation of different models that fit well to specific period and therefore provide the best results.

\subsection{Method}
\begin{figure}[htbp]
\caption{Pipeline of different scenarios}
\includegraphics[width=0.5\textwidth]{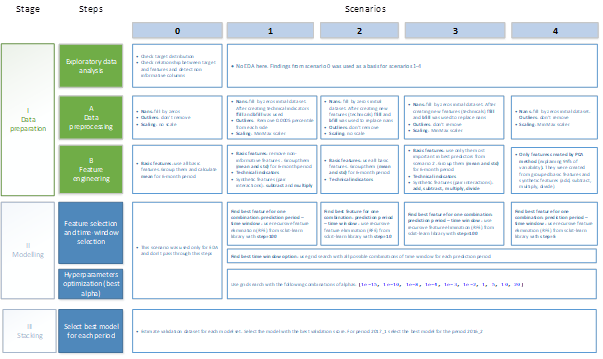}
\label{fig:kirill_pipeline}
\end{figure}

My final solution consists of four sets of linear models (four scenarios) that were built using the classical data science pipeline (see Fig.~\ref{fig:kirill_pipeline}).

The main difference between the scenarios - is the feature engineering step. I generated and tested four group of features:

\begin{table}[htbp]
\caption{Feature types}
\begin{center}
\begin{tabular}{>{\centering\arraybackslash}p{3.4cm}>{\centering\arraybackslash}p{4.4cm}}
\hline
\textbf{A. Grouped basic features} & \textbf{B. Technical indicators} \\
\hline
\begin{tabular}{@{\labelitemi\hspace{\dimexpr\tabcolsep}}p{3.2cm}}In the initial dataset we have anonymized features for each month.\\
As we have to predict 6-month return, these basic features were aggregated\\
I used mean of the basic features and std\\
In absolute and percentage value\end{tabular} &
\begin{tabular}{@{\labelitemi\hspace{\tabcolsep}}p{4.2cm}}Main idea: target value of current period could be basic feature of the next period (e.g predicted return for second half 2002 is the target for the first half of 2002 but feature for second half)\\However, we couldn’t link it with the specific security. All we can do, is to calculate mean value for the whole period. Additionally, we can calculate aggregated technical indicators based on this value: Moving Average, Exponential Moving Average, Momentum, Rate of Change (see full list in paper)\\	Finally, we can “encode” each row in the period by these aggregated values\end{tabular}\\
\hline
\textbf{C. Synthetic features} & \textbf{D. PCA} \\
\hline
\begin{tabular}{@{\labelitemi\hspace{\dimexpr\tabcolsep}}>{\raggedright\arraybackslash}p{3.2cm}}As a basis I used grouped basic features (mean)\\Then, for each combination of feature 1, feature 2 from this subset I generate new features:\\Depending on scenario, generate different combination of synthetic features\end{tabular} &
\begin{tabular}{@{\labelitemi\hspace{\dimexpr\tabcolsep}}p{4.2cm}}At the first step I generated synthetic features\\Then, with python-sklearn, I generated the components that explains at least 99\% of variance and use them as a feature\end{tabular}
\end{tabular}
\label{tab_kirill_1}
\end{center}
\end{table}

Modeling stage was quite common for all scenarios and consisted of the following steps:

\begin{enumerate}
\item Take all features from scenario 1
\item Train models with different combination of features for each prediction period and check the results on validation dataset. Then, select the model with the best Spearman score on validation dataset 
\item Train  these models with different regularization parameters (alpha) and select the models with the best score on validation dataset for each prediction period
\item Repeat step 2-3 for all scenarios
\item Find the best model from all scenarios for each prediction period based on best score in validation dataset. For the period 2017\_1 the best model for period 2016\_2 was used
\end{enumerate}

Fig.~\ref{fig:kirill_algorithm} shows the described algorithm:

\begin{figure}[htbp]
\caption{Algorithm for the modeling stage}
\includegraphics[width=0.5\textwidth]{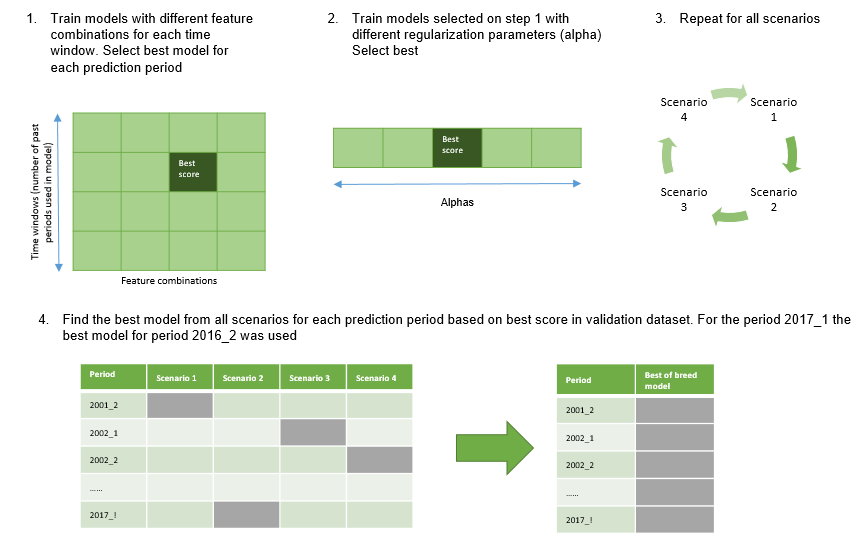}
\label{fig:kirill_algorithm}
\end{figure}

\subsection{Experiments and results}
After training the models with the method describe above, I analyzed the best models and got some interesting findings:

\begin{enumerate}
\item There are small number of top performer features in the dataset. Features $X17$ and $X58$ should be analyzed carefully by company analysts. Also features $X2, X1, X7$ work great in combination, as seen in Fig.~\ref{fig:kirill_top_performer}.

\begin{figure}[htbp]
\caption{Top performer features}
\includegraphics[width=0.5\textwidth]{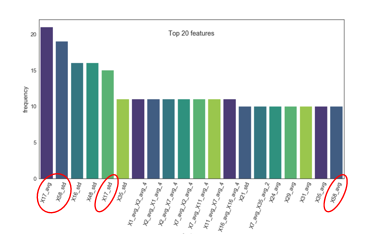}
\label{fig:kirill_top_performer}
\end{figure}

\item Aggregation of basic features, usage of synthetic features and application of dimensionality reduction techniques (PCA) improve the predictive models. Application of technical analysis don’t help when we cannot catch the dynamic of single securities, as we can see in Fig.~\ref{fig:kirill_number_features}.

\begin{figure}[htbp]
\caption{Total number of features by category}
\includegraphics[width=0.5\textwidth]{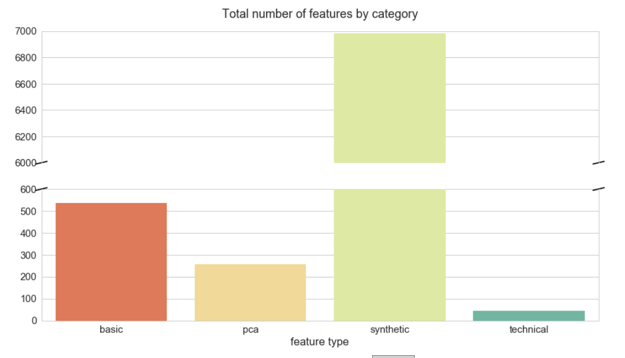}
\label{fig:kirill_number_features}
\end{figure}

\item There is no clear evidence what time windows period is the best for prediction. Perhaps it is the result of current validation technique: 40\% of data was missing inside of each prediction period (Fig.~\ref{fig:kirill_windows_size}).

\begin{figure}[htbp]
\caption{Windows size depending on prediction period}
\includegraphics[width=0.5\textwidth]{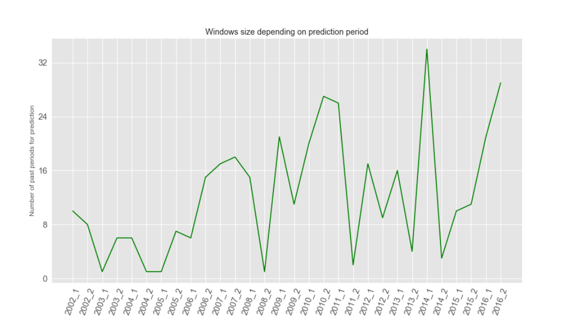}
\label{fig:kirill_windows_size}
\end{figure}

\end{enumerate}

\subsection{Discussion}
During the competition was developed a rigorous pipeline for selecting, testing and evaluating the models. As a result this approach led to top 5 solution during the round 1 and top 4 in final score. 
However there could be some steps to improve the model:

\begin{enumerate}
\item De-anonymize stock dynamics. The stock itself can be anonymize but if we could see the same code for stock x in different period, we could include sequence models in our modelling process and therefore improve model quality
\item Include holdout data that used to score model and retrain model for the future periods. Actually we miss 40\% of data for each of prediction periods and it was a problem to predict future periods
\end{enumerate}

\section{Liu Part}
Wei-Kai Liu
\\

Identifying stocks with the extremely positive and negative performance is of great importance to people ranging from individual to institutional investors. Based on the data of normalized return and multiple stock-level attributes, with feature engineering and feature selection process, machine learning models are built to predict the true ranks of stocks in portfolio, which helps attain the optimal return.

Since the final model aims to be applied on the unseen data in the future, to this end, the approach here is to build a single model framework from linear machine learning model field to cope with the prediction tasks for different time periods, which can yield a more generalized and persistent way to solve this task. After comparison of the models using metrics including Spearman correlation and NDCG based on the true ranks and predicted ranks of stocks for each time period from 2002 to 2016, the best model is a Linear Support Vector Machine Regression model.

\subsection{Method}
Method

\subsubsection{Feature Engineering}
In this task, besides taking the original 420 stock-level attributes into account, new features were also constructed employing transformation and feature engineering. For the newly created features, please refer to Table~\ref{tab_liu_1}.

\begin{table}[htbp]
\caption{Newly Created Features}
\begin{center}
\begin{tabularx}{\columnwidth}{>{\hsize=0.9\hsize\linewidth=\hsize}Y>{\hsize=1.1\hsize\linewidth=\hsize}X}
\hline
\textbf{Feature} & \textbf{Explanation} \\
\hline
X1\texttildelow X70\_mean\_all\_periods & Mean of each original feature from 6 months \\
X1\texttildelow X70\_median\_all\_periods & Median of each original feature from 6 months \\
X1\texttildelow X70\_std\_all\_periods  & Standard deviation of each original feature from 6 months \\
X1\texttildelow X70\_max\_all\_periods  & Maximum of each original feature from 6 months \\
X1\texttildelow X70\_min\_all\_periods  & Minimum of each original feature from 6 months \\
X1\texttildelow X70\_change\_all\_periods & Change from the first value to the last value of each original feature \\
X1\texttildelow X70\_change\_ second\_last\_to\_last\_all\_periods & Change from the second-last value to the last value of each original feature \\
X1\texttildelow X70\_range\_all\_periods & Range of each original feature from 6 months \\
X1\texttildelow X70\_mean\_diff\_all\_periods & Mean of differences of each original feature from 6 months \\
X1\texttildelow X70\_median\_diff\_all\_periods & Median of differences of each original feature from 6 months \\
X1\texttildelow X70\_std\_diff\_all\_periods & Standard deviation of differences of each original feature from 6 months \\
X1\texttildelow X70\_max\_diff\_all\_periods & Maximum of differences of each original feature from 6 months \\
X1\texttildelow X70\_min\_diff\_all\_periods & Minimum of differences of each original feature from 6 months \\
\hline
\end{tabularx}
\label{tab_liu_1}
\end{center}
\end{table}

\subsubsection{Training Framework Setup and Feature Selection}
First, since the approach was to use single model framework to build models on different history data groups and make predictions on various time periods, the relationship between features and target variable, which is the ''Norm\_Ret\_F6M'' column here, for each training and validation time periods group should be considered comprehensively. The validation time periods start from 2002 to 2016. Under the aforementioned premise, whether to take all the history data before validation time period as training data or just regard certain length of the time periods as training data also played an important role.

Second, to counter the problems of overfitting, selecting the optimal feature set for each model rather than all of the features created is a necessary process. Here, the performance of single feature model and correlations within features would serve as the decision-making basis.

\paragraph{Relationship within features and target variable for each training and prediction time periods group and determination on the length of training time periods}
For each feature, Pearson correlation between that and target variable was calculated using multiple length of time periods ranging from 2 to 30 and all the periods, which represents the number of latest time periods before the validation time periods.

Then the Pearson correlation between each feature and target variable was calculated using validation time periods.

After the previous two calculation, the number of different signs of correlation within target variable and each feature under each training and validation time periods group was counted. Each feature was then ranked according to the number of different signs of correlation with target variable, from the least to the largest.

In order to find the best setting for building models, for each length of time periods (from 2 to 30 and all the periods), simple Linear Regression model was built using each of the top 20 features according to the number of different signs of correlation with target variable. Finally, after comparison based on average of Spearman correlation and NDCG, using the most recent 10 periods before validation time periods to get the relationship between features and target variable was suggested. And for training setting, approaching the task using all the history data before validation time period as training data was a better choice.

\paragraph{Selecting optimal feature sets for each model}
After setting up the training framework, features were first ranked according to the number of different signs of correlation with target variable and then features with more than 10 different signs of correlation with target variable were removed.

For the remaining features, Models with only one feature for every feature and model were built, the models used were the same as indicated in the next part. Under each model, features with performance based on average of Spearman correlation and NDCG less than 0 were first gotten rid of and then were ranked from the best to the worst. 

Next, Pearson correlations within each remaining feature were calculated using the average value for each training time period. Finally, after removing the features with Pearson correlation equal to or larger than 0.8 with at least one feature starting from the least significant features, temporarily optimal feature set for each model were generated.

\subsubsection{Modeling}
For modeling, multiple linear machine learning models were applied on the features picked from the previous step. To optimize the performance, different numbers of top features were further chosen and measure by their performance according to the average of Spearman correlation and NDCG from each validation time period. 

The best results of individual model would be displayed at the “RESULTS” part. For the models tested, please refer to Table~\ref{tab_liu_2}.

\begin{table}[htbp]
\caption{Models Used in The Modeling Phase}
\begin{center}
\begin{tabular}{ l c }
\hline
\textbf{Model} & \textbf{Reference} \\
\hline
Linear Regression & \cite{noauthor_sklearn.linear_model.linearregression_nodate} \\
Ridge Regression & \cite{noauthor_sklearn.linear_model.ridge_nodate} \\
Ridge Regression with built-in cross-validation & \cite{noauthor_3.2.4.1.9._nodate} \\
Bayesian Ridge Regression & \cite{noauthor_sklearn.linear_model.bayesianridge_nodate} \\
Huber Regressor & \cite{noauthor_sklearn.linear_model.huberregressor_nodate} \\
Linear Support Vector Machine Regression & \cite{noauthor_sklearn.svm.linearsvr_nodate} \\
\hline
\end{tabular}
\label{tab_liu_2}
\end{center}
\end{table}

\subsection{Experiments and results}

\subsubsection{Comparison of Performance from Each Model with Optimal Feature Set}
After further selecting the best feature set and some settings including hyperparameters for each model, the best model was Linear Support Vector Machine Regression with the top 26 features and with NAs imputed using 0. The average of Spearman correlation and NDCG for each prediction period ranging from 2002 to 2016 is 0.1045. The results for each model are displayed in Table~\ref{tab_liu_3}:

\begin{table}[htbp]
\caption{Performance of Each Model}
\begin{center}
\renewcommand{\tabularxcolumn}[1]{>{\arraybackslash}m{#1}}
\begin{tabularx}{\columnwidth}{>{\hsize=1.6\hsize\linewidth=\hsize}X>{\hsize=0.6\hsize\linewidth=\hsize}Z>{\hsize=0.7\hsize\linewidth=\hsize}Z>{\hsize=1.1\hsize\linewidth=\hsize}Z}
\hline
\textbf{Model} & \textbf{Number of Top Features} & \textbf{Imputation} & \textbf{Performance (Spearman correlation and NDCG avg.)} \\
\hline
Linear Regression & 26 & Median & 0.0999 \\
Ridge Regression & 26 & Median & 0.0985 \\
Ridge Regression with built-in cross-validation & 26 & Median & 0.0957 \\
Bayesian Ridge Regression & 26 & Median & 0.0933 \\
Huber Regressor & 24 & Median & 0.0996 \\
Linear Support Vector Machine Regression & 26 & 0 & 0.1045 \\
\hline
\end{tabularx}
\label{tab_liu_3}
\end{center}
\end{table}

According to the Table~\ref{tab_liu_3}, the best model setting was then applied to train multiple models, which served as models to make final predictions on data from 2002 to 2017 with “Train” column indicating 0.

\subsection{Discussion}
The approach proposed in this section can be summarized by the following four parts, (1) constructing features which considering the characteristics of financial field, (2) framing the proper training setting, (3) selecting features with meaningful predictive capabilities and (4) using single model setting with linear models.

The result above indicates that there is actually some space for improvement. Considering the complexity behind finance-related prediction tasks, for future work, generating more features considering something like interaction based on more detailed observations from exploratory data analysis (EDA) and implementing more complex models including tree-based models and neural network with employing ensembling techniques to address problems of overfitting would form good places to start trying.

\section{Look, Learn and Trade}
Pranoot Hatwar
\\

Stock Market being one of the most volatile field
known to mankind, it has always been a really challenging
task to predict the stock prices. In this
work using deep learning techniques we address the problem
to identify the best performing stocks in each time period using
the provided data sets of financial predictors and semi-annual
returns.

\begin{figure}[htbp]
\caption{Block diagram for the framework}
\includegraphics[width=0.5\textwidth]{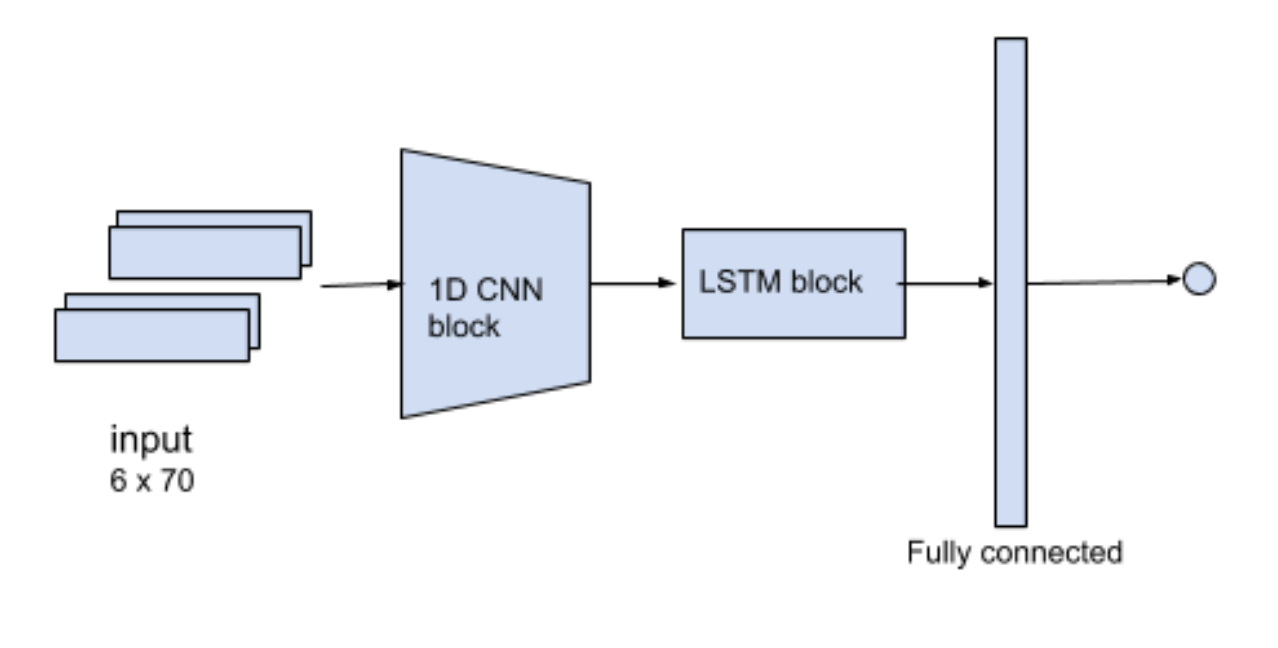}
\label{fig:pranoot_block}
\end{figure}

The framework used have four parts and we separate
the description of framework accordingly. Initally we pre-
process the data by imputing the NA values and then
reshaping it into a sequence of 6 time steps. Second part
consist of convolutional layers, third one consists of recurrent
layer and these two blocks are the main building block of
the framework. The fourth and final block is fully connected
layer which is used for the regression task to predict the
returns in each time period of six months. The framework can be seen in Fig. ~\ref{fig:pranoot_block}.

\subsection{Method}

\subsubsection{Pre Processing}
The input of the framework is a sequence of 70 attributes
over a span of 6 months. So, before putting the data into the
model we impute the NA values with zeros and then reshape
the attributes into 1 {x} 6 {x} 70.

\subsubsection{Convolutional Layers}
The role of convolution layers in the framework is to
extract higher dimensional features for every time step which
can be used by the recurrent layers. 1-D Convolutional layers
were used for learning the higher level representations of the
input sequence. There are in all 70 attributes over period of
6 months which is input to this Convolutional layers. The
output of this block is a sequence of 6 time steps again but
with more richer and useful features for recurrent block.

\subsubsection{Recurrent Layers}
The major reason for using this block was to exploit the
most important property of recurrent neural networks which is to learn
temporal information of a time dependent data. The output
of convolutional block is a sequence of feature vector over
period of 6 months. In this work long short term memory
cells with ’tanh’ activation were used for this block.

\subsubsection{Output Layers}
The embedding generated by the recurrent layers is now fed to a fully connected layer which will use this rich representation of the input sequence for predicting the retun in 6 months forward window. We use
\textit{tanh} activation for the neurons. Training was done with
presence of dropout layer after the recurrent block and to
make the model more robust.

\subsection{Experiments and results}
In this paper we implement a deep learning frame-
work to predict a return over a forward 6-month window using k-fold cross validation. This framework achieved Spearman Correleation of 0.155 and NDCG of 0.166 for the first round.

\subsection{Discussion}
This method is an end-to-end scheme for predicting semi annual returns. The proposed system directly learns
a mapping from the past year data to latent space which helps in predicitng the returns. Future works can be done on the same. 

\begin{enumerate}
\item We can use the sentiment feature vector generated using latest news scraped from the blogs on the internet of the stock for prediction of the returns
\item The imputed values in the pre-processing can be replaced by mean or median of the past data.
\end{enumerate}

\section{crowdAI Challenge}
\begin{figure}[htbp]
\caption{Distribution of normalized stock returns for time period $1996\_2$}
\includegraphics[width=0.5\textwidth]{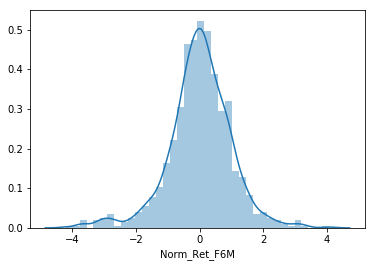}
\label{fig:dist_returns}
\end{figure}

\begin{figure*}
    \centering
    \subfigure[]{\includegraphics[width=0.25\textwidth]{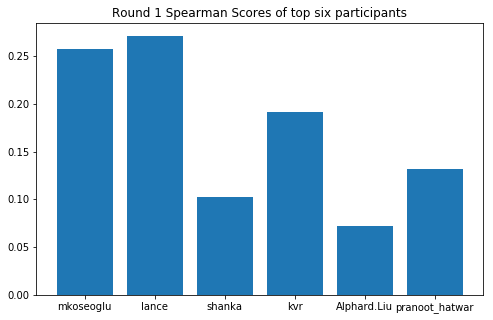}\label{sp1}}%
    \subfigure[]{\includegraphics[width=0.25\textwidth]{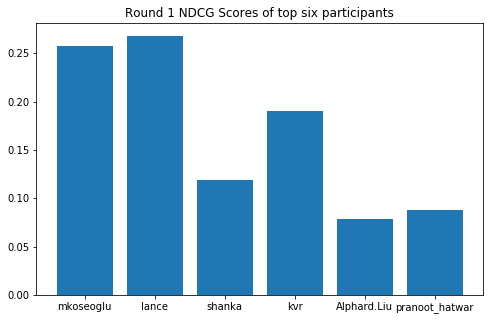}\label{ndcg1}}%
    \subfigure[]{\includegraphics[width=0.25\textwidth]{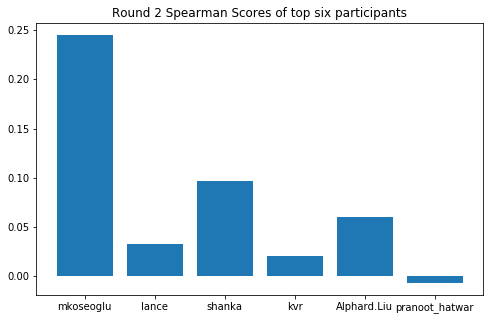}\label{sp2}}%
    \subfigure[]{\includegraphics[width=0.25\textwidth]{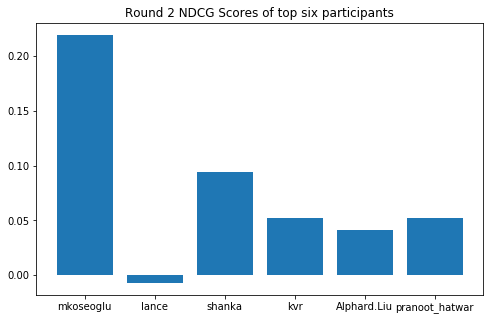}\label{ndcg2}}%
    \caption{Distribution of scores for top six participants in order for Round 1 and Round 2}
    \label{final_scores}
\end{figure*}

\begin{figure}[htbp]
\caption{Total submissions made by the top six participants}
\includegraphics[width=0.5\textwidth]{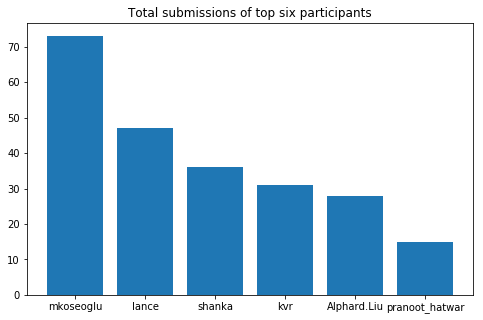}
\label{fig:total_subs}
\end{figure}

\begin{figure}[htbp]
\caption{Distribution of final scores for top six partcipants over Round 1 and Round 2}
\includegraphics[width=0.5\textwidth]{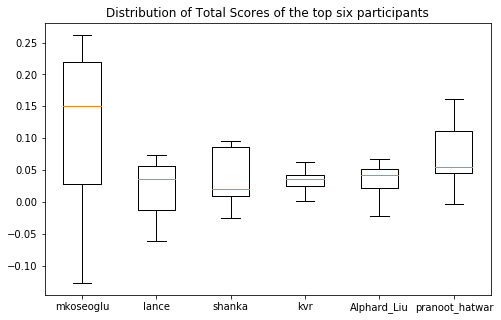}
\label{fig:box_plot}
\end{figure}

\begin{figure}[htbp]
\caption{Day wise total submissions across all participants}
\includegraphics[width=0.5\textwidth]{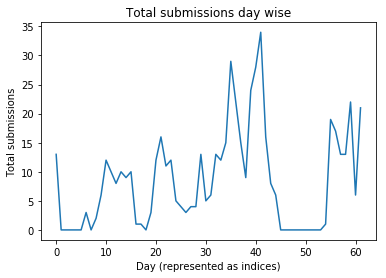}
\label{fig:total_subs_daywise}
\end{figure}

\begin{figure}[htbp]
\caption{Day wise evolution of max spearman correlation over Round 1 and Round 2}
\includegraphics[width=0.5\textwidth]{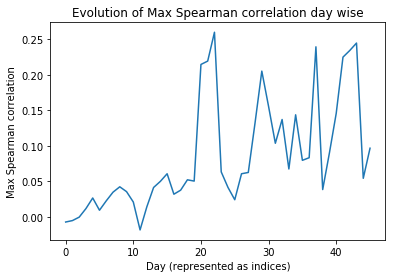}
\label{fig:max_sp}
\end{figure}
The IEEE investment ranking challenge was hosted by crowdAI, where the goal was to build a model that ranks a set of stocks based on the expected return over a forward six month window. The dataset for this task was provided by Principal Financial Group which consisted of predictors and semi-annual returns for a group of stocks from 1996 to 2017. This span of 21 years was represented as 42 non-overlapping 6-month periods. In each of the 42 time periods, roughly 900 stocks with the largest market capitalization (i.e., total market value in USD) were selected. All stock identifiers had been removed and all numeric variables had been anonymized and normalized. Training and test datasets were created by selecting a random sample of stocks at each time period. 60\% of stocks were sampled into the training set and the remaining 40\% created the test set. Finally, all data from the second half of 2017 was allocated to the test set. This 6-month period provided a final out-of-sample test of a model’s performance.
The distribution of the stock returns for the first time period (1996\_2) is shown in Fig. \ref{fig:dist_returns}. 

For the purpose of evaluation an expanding window procedure was chosen. For a given time period, 
$T$, an expanding window test allows the model to incorporate all available information up to time 
$T$, to generate predictions for time 
$T+1$. The metrics used were Spearman correlation and  Normalized Discounted Cumulative Gain of Top 20\% (NDCG).

The challenge was conducted in two rounds , the first round from $20^{th}$ March, 2018 to $3^{rd}$ May, 2018 and second round from from $3^{rd}$ May, 2018 to $20^{th}$ May, 2018. In Round 1, participants were asked to create models and upload predictions to crowdAI and focus on prototyping models that maximized statistical measures on holdout data from 2002-2016. Round 2 was open to all challenge participants. Participants explained their methods, results, and conclusions in short paper and also packaged code of submitted solution using Docker for testing and evaluation on a new set of holdout data from 2017. The final winners were selected based on the final score which was the average of rank of spearman correlation of Round 1, rank of NDCG of Round 1, rank of spearman correlation of Round 2, rank of NDCG of Round 2. The spearman correlation and NDCG for Round 1 for the top six participants from left to right in order of final rank (x axis label of the bar graphs in Fig. \ref{final_scores}) are shown in Fig. \ref{sp1} and Fig. \ref{ndcg1} respectively. For Round 2 evaluation 2017's data was used and the spearman and NDCG scores are shown in Fig. \ref{sp2} and Fig. \ref{ndcg2} respectively.

The distribution showing the total number of submissions made by the top six participants over Round 1 and Round 2 can be found in Fig. \ref{fig:total_subs}, whereas the day wise characteristics of total submissions for all the participants is shown in Fig. \ref{fig:total_subs_daywise}. The box plot in Fig. \ref{fig:box_plot} shows the mean and the variance of the scores calculated as average of NDCG and spearman correlation for the top six participants over Round 1 and Round 2. Finally the evolution of the maximum spearman correlation over both the rounds calculated over the days for which there was at least one submission is shown in Fig. \ref{fig:max_sp}.

\section*{Acknowledgment}
The authors would like to acknowledge Benjamin Harlander, Principal Financial group for providing the challenge data and Ecole Polytechnique Federale de Lausanne (EPFL) especially  Sylvian Bernard and Sean Carroll for help in organizing the competition.

\end{document}